\documentclass[lettersize,journal]{IEEEtran}
\IEEEoverridecommandlockouts
\usepackage{cite}
\usepackage{multirow}
\usepackage{amsmath,amssymb,amsfonts}
\usepackage{algorithmic}
\usepackage{graphicx}
\usepackage{textcomp}
\usepackage{xcolor}
\usepackage{textcomp}
\usepackage{stfloats}
\usepackage{url}
\usepackage{verbatim}
\usepackage{algorithm}
\usepackage{array}
\usepackage[caption=false,font=normalsize,labelfont=sf,textfont=sf]{subfig}
\def\BibTeX{{\rm B\kern-.05em{\sc i\kern-.025em b}\kern-.08em
    T\kern-.1667em\lower.7ex\hbox{E}\kern-.125emX}}
\begin{document}

\title{Diffusion Model-Based Size Variable Virtual Try-On Technology and Evaluation Method \\
{\footnotesize \textsuperscript{*}}
\thanks{}
}

\author{
    Shufang Zhang,~\text{\IEEEmembership{Member,~IEEE}}, 
    Hang Qian, Minxue Ni, Yaxuan Li,Wenxin Ding, and Jun Liu
}

\maketitle
\begin{abstract}
With the rapid development of e-commerce, virtual try-on technology has become an essential tool to satisfy consumers' personalized clothing preferences. Diffusion-based virtual try-on systems aim to naturally align garments with target individuals, generating realistic and detailed try-on images. However, existing methods overlook the importance of garment size variations in meeting personalized consumer needs. To address this, we propose a novel virtual try-on method named SV-VTON, which introduces garment sizing concepts into virtual try-on tasks. The SV-VTON method first generates refined masks for multiple garment sizes, then integrates these masks with garment images at varying proportions, enabling virtual try-on simulations across different sizes. In addition, we developed a specialized size evaluation module to quantitatively assess the accuracy of size variations. This module calculates differences between generated size increments and international sizing standards, providing objective measurements of size accuracy. To further validate SV-VTON's generalization capability across different models, we conducted experiments on multiple SOTA Diffusion models. The results demonstrate that SV-VTON consistently achieves precise multi-size virtual try-on across various SOTA models, and validates the effectiveness and rationality of the proposed method, significantly fulfilling users' personalized multi-size virtual try-on requirements.
\end{abstract}


\begin{figure}[t]
\centerline{\includegraphics[width=0.8\columnwidth]{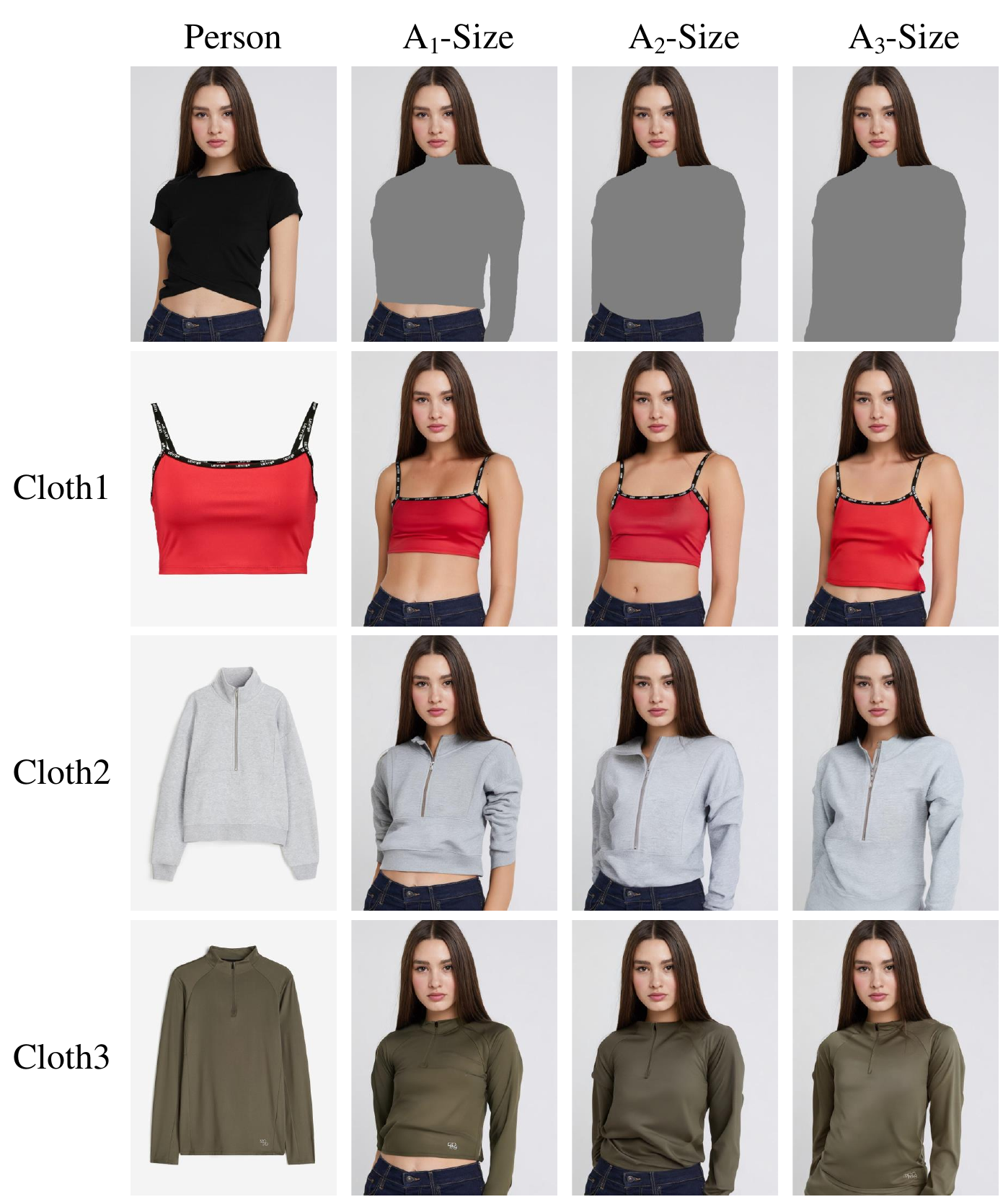}}
\caption{Multi-size VTON results generated by our method. Our approach can generate images for the same person in different garment sizes.}
\label{fig1}
\end{figure}

\section{Introduction}

As online commerce grows rapidly, virtual try-on (VTON) technology is increasingly critical in online shopping. VTON methods aim to progressively narrow the gap between generated images and real garments in terms of size and texture details. Early VTON approaches primarily employed TPS (Thin-Plate Spline) transformations \cite{1,7,10} and appearance flow techniques \cite{2,18,21} to address garment distortion and warping problems. Recent studies have integrated advanced human segmentation and pose estimation methods \cite{3,4,6,8} to optimize garment warping processes. However, with the advent of Diffusion model \cite{24}, the powerful generative capabilities have significantly improved garment warping handling. Consequently, as a crucial input condition for Diffusion model, the degree of garment warping is primarily determined by the shape of the mask.

However, existing methods typically utilize a single tight or loose mask for model training. As illustrated in Fig.~2, models trained with tight masks tend to generate fitting results closely aligned with the mask shape, whereas those trained with loose masks produce outputs closer to the original garment length. Such a single-fitting outcome is insufficient to meet consumers' diverse sizing preferences (tight, fitted, or loose styles). Additionally, most existing VTON methods employ only a single mask and corresponding garment data during training and testing, causing the generated garments to be poorly generalized across diverse body shapes \cite{9,10,11,12}. Consequently, ignoring implicit size information causes models to produce garments that are predominantly limited to a single-fitting style.

\begin{figure}[t]
\centerline{\includegraphics[width=0.8\columnwidth]{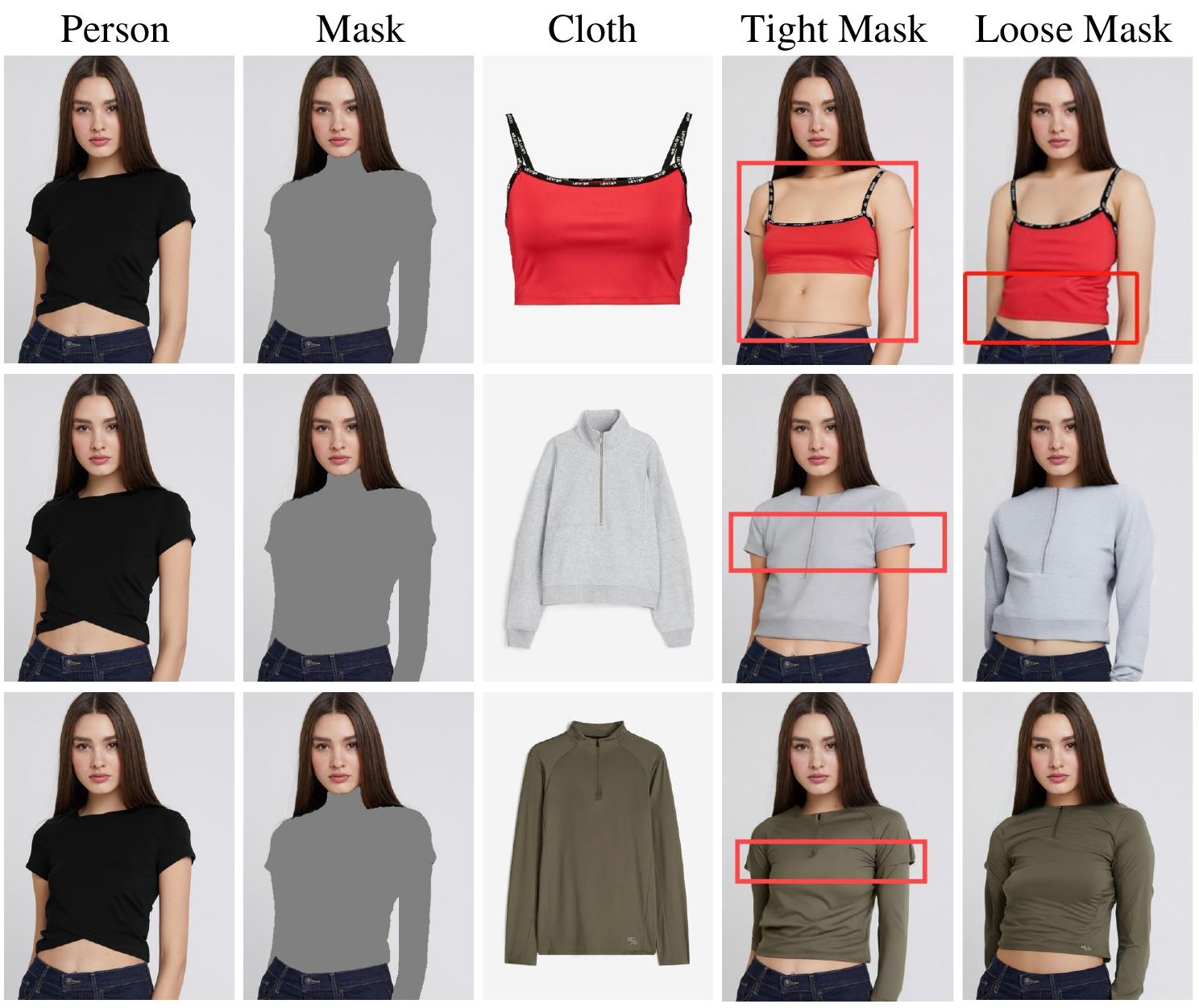}}
\caption{Testing results using models trained with loose masks (right) and tight masks (left). The regions marked by red boxes indicate problematic areas caused by single-size mask generation.}
\label{fig2}
\end{figure}

This paper proposes a novel approach named Size Variation Virtual Try On(SV-VTON) to address these issues. Specifically, the method generates refined masks for multiple garment sizes, and adjusts the proportions of the garment image to simulate diverse fitting conditions, thereby realizing generated effects for tight ($A_1$), fitted ($A_2$), and loose ($A_3$) styles (as illustrated in Fig.~1). SV-VTON first generates coarse initial masks based on human keypoints and then refines mask edges for smoothness and accuracy through an Edge Attention mechanism combined with the U$^2$-Net model \cite{48}. To mitigate inherent dependencies on original garment images and mask shapes, SV-VTON adopts a multi-size mask fusion strategy during training. 
Furthermore, considering that international standard size increments typically have fixed values, an independent Evaluation Module (EM) is designed to verify sizing accuracy quantitatively. This module measures the differences between generated garment sizing increments and international standard increments. Both qualitative and quantitative experiments demonstrate that SV-VTON significantly improves the accuracy and diversity of garment fitting outcomes across various body sizes, effectively meeting users' personalized preferences. Quantitative analyses further reveal sizing errors of only approximately 5\%-10\%.

The proposed method framework offers three key advantages:
\begin{enumerate}
    \item \textbf{Multi-size VTON Capability:} SV-VTON simulates multiple sizing conditions to achieve diverse virtual fitting styles, specifically covering tight ($A_1$), fitted ($A_2$), and loose ($A_3$) garment fits, thus obtaining a more satisfying fitting experience.
    \item \textbf{Two-stage Refined Multi-size Mask Generation Strategy:} We first generate a coarse mask based on human keypoints and subsequently refine the mask's edges using an Edge Attention mechanism combined with the U$^2$-Net model, significantly improving the accuracy of sizing variations.
    \item \textbf{Novel Quantitative Evaluation Metric for Size Variation:} This paper introduces a novel quantitative evaluation metric for size variation, comparing the size increments of multi-size generated results with standardized international size increments across four key dimensions.
\end{enumerate}

\section{Related Work}

\subsection{Image-based Virtual Try-On}

In previous studies, most VTON methods GAN as frameworks~\cite{7,18,19,20}, typically dividing the task into two independent modules: (1) a clothing warping module, responsible for warping garments to the designated regions of target persons; and (2) a GAN-based image generation module, responsible for blending the warped garments with target images to generate the final generated results. To achieve more accurate clothing warping, existing methods~\cite{16,17,18,21} commonly utilize neural networks to predict appearance flow maps~\cite{2,8,11}. Additionally, various improvement strategies have been proposed to address issues related to insufficient garment fitting with target bodies after warping~\cite{7,13,14,15}.

Recently, Diffusion models have demonstrated significant advantages over GAN-based approaches in image generation tasks. Many researchers have shifted their focus toward virtual try-on tasks using large-scale pre-trained Diffusion models~\cite{5,22,25,26,28}. For instance, Reference Net ~\cite{5} introduces a spatial attention mechanism to encode features from the image into the Denoising UNet, significantly enhancing the texture and detail of the virtual try-on results. Stable Diffusion ~\cite{27,45} exhibits outstanding performance in virtual try-on tasks due to its flexible mask inpainting and text-guidance capabilities. DiffusionCLIP~\cite{31} combines pre-trained Diffusion models with CLIP loss to effectively refine image details. Several studies have further introduced precise control mechanisms for generating images by adjusting input conditions~\cite{13,32,33,34,42}. For example, DCI-VTON~\cite{29} utilizes garment images processed by a warping module as inputs to Diffusion models, effectively controlling the try-on outcomes. However, existing methods often neglect the realistic sizing relationship between generated garments and human bodies in multi-size virtual try-on tasks, resulting in garments either overly conforming to the target person's body or excessively relying on the original mask shapes. To overcome these limitations, this paper introduces refined multi-size masks to assist model training, freeing the model from inherent dependencies on original garment lengths and mask shapes, as illustrated in Fig.~1, thus achieving more accurate multi-size generated effects.

\subsection{Virtual Try-On Evaluation Metrics}

Early quantitative evaluation metrics in virtual try-on tasks mainly focused on image quality, employing pixel-level similarity measurements between generated and real images. Pixel-level evaluation methods, such as SSIM and PSNR, have been widely used in earlier studies~\cite{7,18,19,20,38} to quantify generation quality. Subsequently, FID~\cite{49}and KID~\cite{50} quantitatively characterize the similarity between the generated results and test dataset distributions—reflecting overall distribution consistency. Multi-Scale Structural Similarity (MS-SSIM)~\cite{51} is capable of evaluating the structural similarity of images across multiple scales. LPIPS~\cite{39} calculates the distance in the feature space instead of pixel-level differences. CLIPScore~\cite{44} utilizes the CLIP model to measure the images and generated descriptions' similarity. In recent years, evaluation methods have evolved towards more comprehensive semantic-level assessments, emphasizing the semantic structural rationality and accuracy of garments~\cite{19,40,41,47}.

Additionally, numerous no-reference image quality assessment metrics have been proposed in recent years. Multi-dimension Attention Network for no-reference Image Quality Assessment (MANIQA)~\cite{52} is a no-reference method that evaluates image quality without needing a comparison with real images, leveraging a multi-dimensional attention mechanism to comprehensively capture image clarity and realism. Multi-Scale Image Quality Transformer (MUSIQ)~\cite{53} is a no-reference evaluation method based on a Transformer architecture, which effectively captures multi-scale features to perform both global and local image quality assessments. 

\section{Methodology}

The overall workflow of the proposed SV-VTON and EM is illustrated in Fig.~3. SV-VTON includes Multi-size Mask Generation Module (MMGM) and Try On Module (TOM). MMGM takes \( M_o \) as input and generates multi-size refined masks as output. Then, TOM processes the multi-size masks along with the corresponding proportionally adjusted garments to produce multi-size virtual try-on images.

Moreover, we propose an independent EM to quantitatively assess the effectiveness and validity of size variations in the generated images. As shown in Fig.~3, EM evaluates sizing accuracy by systematically comparing size increments of generated garments against international standard size increments. EM aggregates deviations measured across multiple Diffusion models to derive a comprehensive assessment metric, objectively reflecting both the sizing accuracy and generalization capabilities of the SV-VTON method.

\begin{figure*}[t]
\centerline{\includegraphics[width=\textwidth]{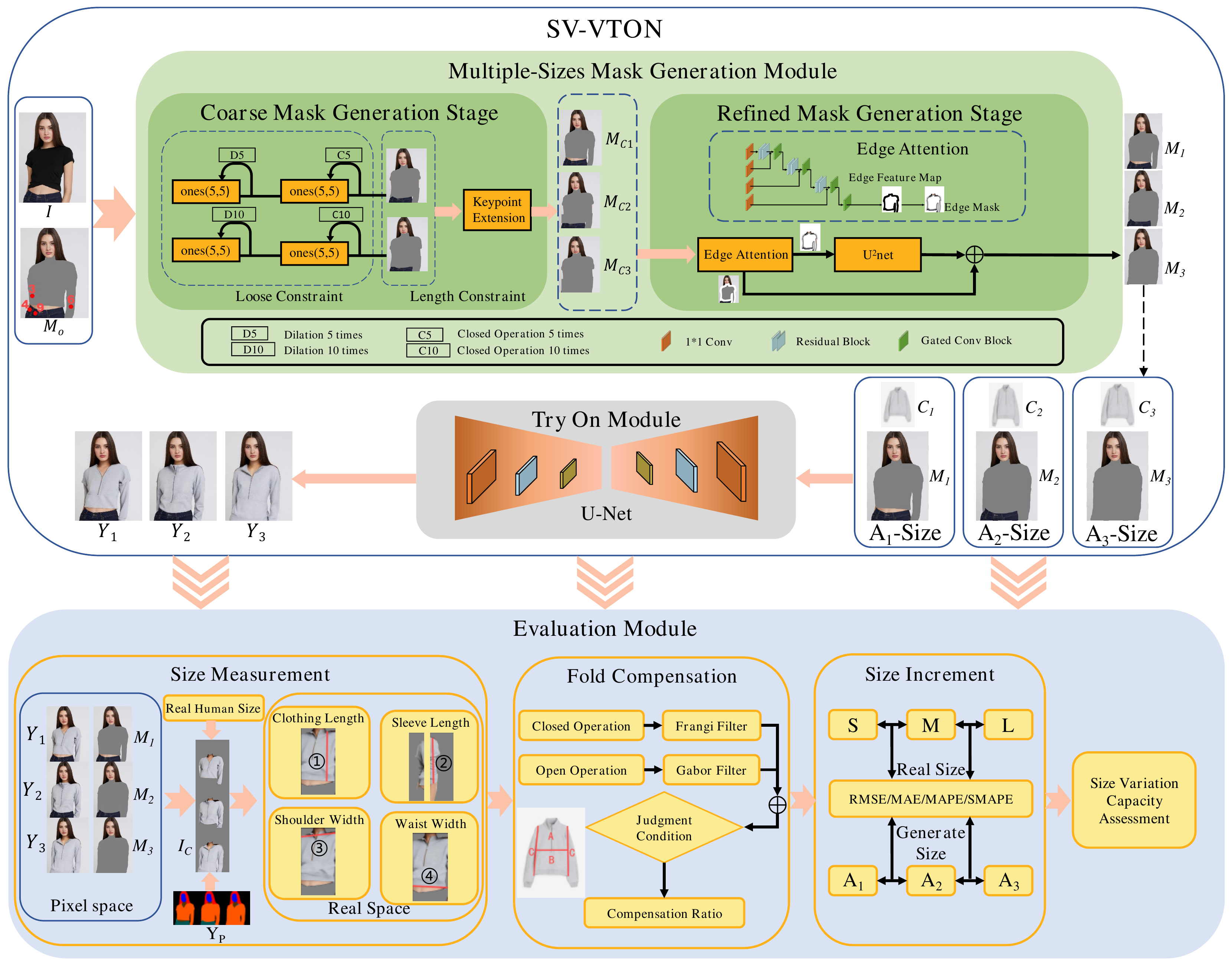}}
\caption{Overview of the proposed framework of SV-VTON and EM. We use $I$ and $M_o$ to denote the human source image and original tight-fitting mask, respectively. The Multiple-Size Mask Generation Module includes two stages: the Coarse Mask Generation Stage and the Refined Mask Generation Stage, which respectively produce a coarse mask $M_C$ and refined masks ($M_1$, $M_2$, and $M_3$) corresponding to $A_1$, $A_2$, and $A_3$ sizes. In the Try-On Module, based on the Diffusion model, the multi-size masks and proportionally adjusted garment images ($C_1$, $C_2$, and $C_3$) are inputted to generate multiple results ($Y_1$, $Y_2$, and $Y_3$) for the same garment under different sizes. The EM quantitatively measures size increments using four defined metrics, considering garment wrinkles as compensation for different size proportions. It computes deviations between the size increments of generated images and international standard increments to comprehensively evaluate the validity and accuracy of garment sizing variations.}
\label{fig3}
\end{figure*}

\subsection{SV-VTON} \label{SV-VTON}
MMGM consists of two stages: the Coarse Mask Generation Stage (CMGS) and the Refined Mask Generation Stage (RMGS). In CMGS, masks of varying looseness and length—corresponding to $A_2$ and $A_3$ sizes—are produced. Then, in RMGS, an edge smoothing strategy refines the coarse edges of masks generated in CMGS, yielding refined and smooth multi-size masks. This two-stage strategy enables SV-VTON to effectively simulate diverse garment size conditions, significantly improving the quality and size accuracy of generated images. 

In TOM, we leverage the powerful generative capability of the Diffusion model, using these refined multi-size masks fused with proportionally adjusted garment images as inputs to generate high-quality virtual try-on images for multiple garment sizes.

\subsubsection{Coarse Mask Generation Stage (CMGS)} \label{CMGS}
In CMGS, we designed a mask refinement strategy to precisely control the generated virtual try-on masks regarding looseness and garment length for different clothing sizes, as illustrated in Equation~(\ref{eq: MC}).

\begin{equation}
M_{C}(\lambda)=
\left\{
\begin{array}{l}
(x,y) \in \phi^{5(\lambda-1)}\Bigl(\delta^{5(\lambda-1)}\bigl(G_{5\times5}(M_{O})\bigr)\Bigr)\\[8pt]
\text{s.t.} \quad x_{4}\leq x\leq x_{4}+|x_{6}-x_{3}|\\[5pt]
\quad\;\;\;\; y_{3}\leq y\leq y_{3}+0.8(\lambda-1)\,|y_{9}-y_{3}|\\[5pt] 
\quad\;\;\;\;1\leq \lambda\leq3
\end{array}
\right\}
\label{eq: MC}
\end{equation}

In Equation~(\ref{eq: MC}), the parameter \( \lambda \) is utilized to control incremental garment sizing. Specifically, when \( \lambda = 1 \), the mask \( M_{O} \) is selected as tight-fitting size \( M_{C1} \). Subsequently, as \( \lambda \) increases to 2 and 3, the garment size expands by one standard sizing interval each time, yielding masks \( M_{C2} \) and \( M_{C3} \), respectively. Here, \( G_{5\times5}(\cdot) \) denotes the Gaussian smoothing operation using a \( 5\times5 \) kernel. The operators \( \delta^{i}(\cdot) \) and \( \phi^{i}(\cdot) \) represent the dilation and closing operations applied iteratively \( i \) times, respectively; dilation iterations control the garment looseness while the closing operations refine the mask boundaries. Subsequently, we implement the length constraint via the keypoint extension method. The extension of the mask's starting position is defined by the horizontal coordinate \( x_4 \) of the right wrist and the vertical coordinate \( y_3 \) of the right elbow. The mask's width is determined by the horizontal distance \( |x_{6} - x_{3}| \) (i.e., the difference between the right and left elbows), and its vertical extent is set as \( 0.8 \) times the distance between the right elbow and right waist, scaled by \( \lambda \). This results in a vertical extension of \( 0.8|y_{9} - y_{3}| \) for \( A_2 \) style and \( 1.6|y_{9} - y_{3}| \) for \( A_3 \) style.

\subsubsection{Refined Mask Generation Stage (RMGS)} \label{RMGS}

Due to the generated masks often exhibiting rough and irregular edges, studies have demonstrated that edge information is a crucial prior constraint and plays a significant role in object segmentation tasks~\cite{36,37}. In the RMGS stage, to effectively address this issue, we introduce an Edge Attention mechanism and U$^2$-Net, which selectively smooths the mask's rough edges without altering the image's background regions. 

Edge Attention Block includes a Convolutional Neural Network (CNN) architecture, as illustrated in Fig.3. This module is specifically designed to extract and refine the edge features of the generated masks. The network architecture comprises four parallel $1\times1$ convolutional layers as the initial feature extraction module. Each of these convolutional layers is followed by a sequence of Residual Blocks and Gated Convolution Blocks to enhance the extraction of rough edge features, while preserving critical edge details. By integrating this Edge Attention mechanism, our method effectively smooths the mask's edges without affecting its structural integrity, thereby improving the accuracy of the generated multi-size virtual try-on results.

The gated convolution block computation process is formulated as follows:
First, the input feature map $F_i$ undergoes a $1\times1$ convolution operation, followed by a sigmoid activation function to generate the gating weight $F_o$. The final refined edge feature map $E_o$ is then obtained by applying the attention map $F_o$ to weight the original edge feature $E_i$, followed by a residual connection and channel-wise weighting with convolution kernel $W$:

\begin{equation}
E_o(i, j) = \Big(E_i(i, j) \odot \sigma(\text{Conv}(F_i)) + E_i(i, j) \Big)^{T} \cdot W,
\label{eq:combined}
\end{equation}

where $\sigma(\cdot)$ represents the sigmoid activation function that constrains the gating values to the range of $[0,1]$, ensuring selective attention to the mask's edge regions, and $\odot$ denotes the element-wise multiplication (Hadamard product). This formulation effectively refines the extracted edge features while preserving essential information via residual connections. The enhanced feature map $E_o$ is then propagated to subsequent layers of the Edge Stream for further refinement. 
The final Edge Feature Map is utilized to construct an Edge Mask, which is subsequently fed into the U$^2$-Net for additional edge smoothing and fine-grained refinement. 

Subsequently, the multi-size masks and proportionally adjusted garment images are inputted to generate multiple virtual try-on results (\( Y_1 \), \( Y_2 \), \( Y_3 \)) for the same garment in different sizes. The scaling factors of the proportionally adjusted garment images will be elaborated in the ablation study later.

\subsection{Evaluation Module} \label{EM}
To accurately assess the effectiveness and reasonableness of size variations in multi-size generated images, we propose an independent EM. This module quantitatively evaluates the precision of generated garment size variations based on multi-dimensional measurement metrics. Given that international size increments are standardized, the EM module compares the sizing increments derived from generated images ($A_1$–$A_2$, $A_2$–$A_3$) and international standardized increments. This approach objectively assesses generated results' accuracy by quantifying deviations from standardized sizing benchmarks.

\subsubsection{Size Measurement}
Since the generated garment area is contained within the input mask region, extracting the target garment region from the refined mask is necessary before performing measurements. The extraction process consists of two steps. First, the generated image and its corresponding refined mask are input into the module to obtain the generated garment region, denoted as $I_C$. At the same time, other areas are covered with masks. Then, the extracted garment region $I_C$ is compared with the original semantic segmentation map $Y_P$ to isolate the garment's central body region and sleeve region. Subsequently, the four key measurement dimensions—CL, SL, SW, and WW—are defined and computed separately within the extracted body and sleeve regions, as illustrated in Fig.~4. Finally, based on the correspondence between real and generated models, we construct a mapping from the generated images' pixel space to real-world physical dimensions, thereby obtaining accurate garment measurements.  

\begin{figure}[t]
\centerline{\includegraphics[width=0.95\columnwidth]{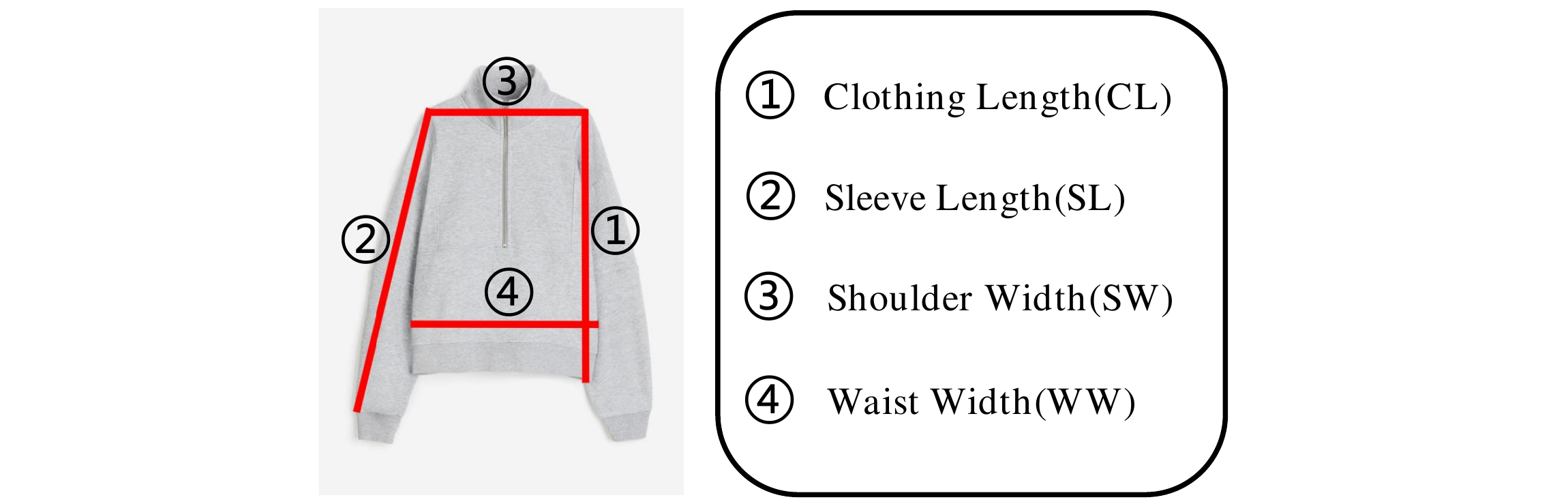}}
\caption{Defined measurement standards. The lengths of the four red lines represent the measured dimensions, indicating the defined measurement positions, which are consistent with international size definitions.}
\label{fig:measurement_standards}
\end{figure}

\subsubsection{Wrinkle Compensation Mechanism}
In the generated images, garments often exhibit wrinkles, which can cause initial size measurements to be underestimated. As a result, the actual visual differences between different sizes may become less distinguishable. To correct this measurement error, the EM introduces a wrinkle-based size compensation mechanism to enhance measurement accuracy.

Specifically, for each measurement dimension, we perform operations along two branches—Frangi filtering after a closing operation, and multi-angle averaged Gabor filtering after an opening operation. The results from these branches are then fused through weighted integration to extract effective wrinkle features. The extracted features are subsequently used to classify wrinkle types according to the following judgment conditions, as illustrated in Fig.~3:

\begin{itemize}
    \item \textbf{Cloth Length Wrinkles:} Wrinkles oriented at $0^\circ$–$45^\circ$ in the region A and region B.
    \item \textbf{Shoulder Width Wrinkles:} Wrinkles oriented at $45^\circ$–$90^\circ$ in the region A.
    \item \textbf{Waist Width Wrinkles:} Wrinkles oriented at $45^\circ$–$90^\circ$ in the region B.
    \item \textbf{Sleeve Length Wrinkles:} Wrinkles oriented at $0^\circ$–$45^\circ$ in the region C.
\end{itemize}

 Finally, the total wrinkle length $L$ is computed by extracting the skeleton length of the wrinkle regions. Based on multiple experimental results, we establish the relationship between the size compensation ratio $R(L)$ and the total wrinkle length $L$, as formulated in Equation~(\ref{eq: wrinkle_compensation}).
\begin{equation}
R(L) =
\begin{cases}
0, & L < A, \\[2mm]
\frac{L-A}{250}\%, & A \le L < B, \\[2mm]
\frac{B-A}{250}\% + \frac{L-B}{200}\%, & B \le L < C, \\[2mm]
\frac{B-A}{250}\% + \frac{C-B}{200}\% + \frac{L-C}{250}\%, & C \le L < D, \\[2mm]
100\%, & D \le L \\[2mm]
\end{cases}
\label{eq: wrinkle_compensation}
\end{equation}

According to extensive experimental analysis, the parameters are set as constants: \( A \), \( B \), \( C \), and \( D \), corresponding to 5000, 10000, 15000, and 20000 pixels, respectively. Using R(L) and L, we can get the length of the actual generated garment region after compensation for wrinkles, as formulated in Equation (4).

\begin{equation}
TL = L + L \cdot R(L)
\label{eq: Ture Length}
\end{equation}

The parameter \( TL \) denotes the length of the actual generated garment region after compensation for wrinkles, as formulated in Equation~(\ref{eq: Ture Length}). This compensation function ensures that garments with more significant wrinkle presence receive a proportionally larger size adjustment, thereby improving the accuracy and realism of multi-size VTON results.

\subsubsection{Size Increment Analysis}

Considering that the actual dimensions of each garment exhibit individual variations, whereas international size increments are standardized, we introduce the Size Increment metric to evaluate the differences between generated garment sizes and international standards from the perspective of size variation. The international size increments for the four measurement metrics are set to approximately 3 cm, 1 cm, 2 cm, and 3 cm, respectively. For computational convenience, we map \( TL \) from pixel space to the real physical space \( PL \), measured in centimetre and denote the clothing length (\( CL_i \)), sleeve length (\( SL_i \)), shoulder width (\( SW_i \)), and waist width (\( WW_i \)) for size \( A_i \) in centimeters, as shown in Equation~(\ref{eq: SI}). \( CL_{ij} \) denotes the size increment in CL between sizes \( A_i \) and \( A_j \). Similarly, \( SL_{ij} \), \( SW_{ij} \), and \( WW_{ij} \) represent the size increments in SL, SW, and WW, respectively, between sizes \( A_i \) and \( A_j \). Taking sizes \( A_1 \) and \( A_2 \) as an example, \( CL_{12} \) specifically denotes the incremental difference in CL between these two sizes.

\begin{equation}
\begin{cases}
CL_{ij} = \left| CL_i - CL_j \right| \\[2mm]
SL_{ij} = \left| SL_i - SL_j \right| \\[2mm]
SW_{ij} = \left| SW_i - SW_j \right| \\[2mm]
WW_{ij} = \left| WW_i - WW_j \right|
\end{cases}
\label{eq: SI}
\end{equation}

\subsubsection{Overall Evaluation of Size Differentiation Ability} 
To analyze errors, we compare the aforementioned size increments against the corresponding international standard increments, employing four statistical evaluation metrics: Mean Absolute Error (MAE), Root Mean Square Error (RMSE), Mean Absolute Percentage Error (MAPE), and Symmetric Mean Absolute Percentage Error (SMAPE). MAE and RMSE are employed to assess absolute size increment errors, whereas MAPE and SMAPE evaluate relative size increment errors. 
Finally, we perform a weighted summation based on the four measurement dimensions to assess the overall accuracy of size variations across different body regions in SV-VTON-generated images. The weights are determined according to the relative importance of each dimension in real-world physical space, as formulated in Equation~(\ref{eq: size-sensitivity-based weighted scoring}). 

\begin{equation}
\begin{aligned}
X_t = \ X_{CL} \cdot \frac{CL_i}{CL_i + SL_i + SW_i + WW_i} \\[2mm]
+ X_{SL} \cdot \frac{SL_i}{CL_i + SL_i + SW_i + WW_i} \\[2mm]
+ X_{SW} \cdot \frac{SW_i}{CL_i + SL_i + SW_i + WW_i} \\[2mm]
+ X_{WW} \cdot \frac{WW_i}{CL_i + SL_i + SW_i + WW_i} 
\end{aligned}
\label{eq: size-sensitivity-based weighted scoring}
\end{equation}

In Equation~(\ref{eq: size-sensitivity-based weighted scoring}), \( X_{CL} \), \( X_{SL} \), \( X_{SW} \), and \( X_{WW} \)represent the size increment errors for their respective dimensions. After being processed through Equation~(\ref{eq: evaluation}), they yield the corresponding evaluation scores \(E(X_{CL})\), \(E(X_{SL})\), \(E(X_{SW})\), and \(E(X_{WW})\), which are then weighted according to their actual physical significance. The weighted result represents the final comprehensive score, denoted as \( E_t \). This size-sensitivity-based weighted scoring approach effectively reflects users' actual concerns regarding different garment region sizes in practical applications. Through the measurements above and comprehensive scoring, the EM provides an objective and quantitative evaluation framework for multi-size VTON methods, contributing to enhanced realism and user satisfaction in virtual try-on systems.

\section{Experiments}

\subsection{Experimental Setup}

\paragraph{Datasets} 

To validate the effectiveness of the proposed method, we conduct experiments on the VITON-HD dataset. The VITON-HD dataset comprises 13,679 image pairs, each consisting of a half-body human view and corresponding upper-body in-store clothing images across 13 categories, with a resolution of $1024 \times 768$.

\paragraph{Implementation Details} 

The multi-size mask fusion strategy is implemented on a single NVIDIA A6000X GPU with a batch size of 4. We employ the Adam optimizer during training to perform gradient descent optimization, setting the learning rate to $2.0 \times 10^{-5}$. Furthermore, a progressive iterative training strategy is adopted. The model is initially trained on images with a resolution of 192, and the resolution is gradually increased until it reaches 768. This strategy enhances the model's performance on high-resolution images.

\paragraph{Evaluation Metrics} 
Our objective is to apply the target garment's shape and texture to the source person's image while ensuring that the generated garment closely matches the actual size of the target garment, thereby maximizing the realism of the VTON scenario. To comprehensively evaluate the accuracy of the model's size variation, we assess the generated images based on four key dimensions and their corresponding error metrics. EM measures the accuracy of multi-size generated size variations in terms of garment length and looseness across four aspects. We adopt internationally standardized size increments as ground-truth values. Then, we compute the errors between the measured increments and these ground truths for each dimension, utilizing four error metrics—MAE, RMSE, MAPE, and SMAPE. Finally, a weighted summation of these error values is performed to derive a comprehensive score that reflects the effectiveness and reasonableness of the generated size variations.

\subsection{Effectiveness Analysis of SV-VTON}
\subsubsection{Qualitative Comparison}

\begin{figure*}[htbp]
\centerline{\includegraphics[width=0.9\textwidth]{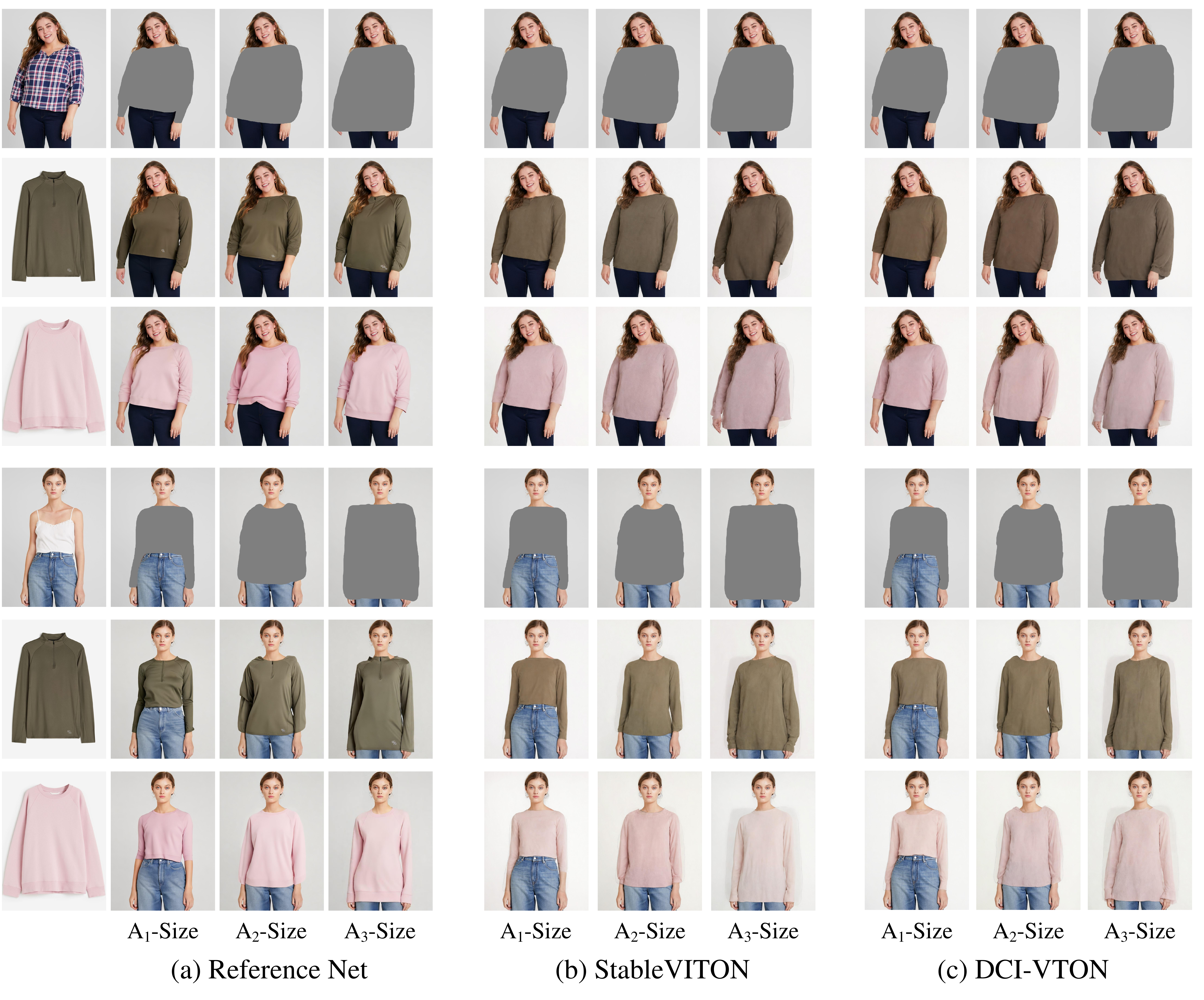}}
\caption{Comparison of generated results across three size styles ($A_1$-size, $A_2$-size, $A_3$-size) using three different Diffusion models: Reference Net, StableVITON, and DCI-VTON.}
\label{fig5}
\end{figure*}

As shown in Fig. 5, the proposed SV-VTON method successfully generates multi-size virtual try-on results while accurately preserving garment warping and texture details. Unlike previous studies that typically consider only the tight-fit size as a standard and fail to meet diverse user demands, SV-VTON can generate virtual try-on results for three size styles: $A_1$, $A_2$, and $A_3$. SV-VTON defaults to the $ A_1$ size for the best fitting. 

Furthermore, to further validate the proposed method's generalization capability, we conduct transfer experiments on three different Diffusion models: Reference Net~\cite{4}, StableVITON~\cite{45}, and DCI-VTON~\cite{29}. The results indicate that SV-VTON maintains precise size variations across different models. However, due to variations in generative capabilities among models, the generated try-on images exhibit subtle differences in wrinkles and detail rendering. Specifically, due to adopting a dual-layer Diffusion model architecture with an integrated spatial attention mechanism, Reference Net significantly enhances the preservation of garment wrinkle details, exhibiting precise size variations in the generated try-on images. StableVITON, which employs a single-layer Diffusion model, maintains precise size variations in the generated virtual try-on images, but exhibits slightly coarse wrinkle generation and moderate variations in garment color. Meanwhile, DCI-VTON utilizes a PF-AFN-based garment warping module to preprocess input garments, which may introduce subtle distortions during multi-size transformations in SV-VTON. While it effectively maintains reasonable size variations in the generated try-on images, some wrinkle details are softened, and garment color shifts more noticeably, slightly impacting texture fidelity.

\subsubsection{Quantitative Comparison}

Based on multi-size try-on images generated by Reference Net, StableVITON, and DCI-VTON, we measured incremental size differences between $A_1$—$A_2$ and $A_2$—$A_3$ across CL, SL, SW, and WW dimensions. We used MAPE and SMAPE metrics to compare the obtained size increments against internationally recognized size standards. As Table~(\ref{tab: increment_error}) indicates, we quantitatively assessed SV-VTON-generated images across the three Diffusion models regarding incremental sizing accuracy. Generally, relative errors below 5\% indicate high measurement accuracy, while errors approaching 5\% remain acceptable. Errors for clothing length remained slightly below 5\%, while those for shoulder and waist widths approached this threshold. This is because the attention mechanism in the generative model primarily focuses on clothing length, resulting in relatively weaker performance in capturing size variations for other dimensions. These results confirm SV-VTON's effectiveness in accurately capturing size variations that are aligned with international sizing standards.
\begin{table}[htbp]
\caption{Increment Error Evaluation (MAPE and SMAPE) for Diffusion Models}
\begin{center}
\renewcommand{\arraystretch}{1.2} 
\setlength{\tabcolsep}{1pt} 
\begin{tabular}{|c|c|cc|cc|cc|}
\hline
\multicolumn{2}{|c|}{\multirow{2}{*}{\textbf{Increment Error}}} 
& \multicolumn{2}{c|}{\textbf{Reference Net}} 
& \multicolumn{2}{c|}{\textbf{StableVITON}} 
& \multicolumn{2}{c|}{\textbf{DCI-VTON}} \\
\cline{3-8}
\multicolumn{2}{|c|}{} 
& \textbf{MAPE}$\downarrow$ & \textbf{SMAPE}$\downarrow$ 
& \textbf{MAPE} & \textbf{SMAPE} 
& \textbf{MAPE} & \textbf{SMAPE} \\
\hline
\multirow{4}{*}{\rotatebox{90}{\textbf{A1-A2}}} & \textbf{Clothing Length} & 3.64 & 3.68 & 3.85 & 3.96 & 4.92 & 5.26 \\
 & \textbf{Sleeve Length}   & 5.50 & 5.90 & 5.30 & 5.76 & 6.48 & 6.52 \\
 & \textbf{Shoulder Width}  & 4.56 & 4.78 & 4.24 & 4.52 & 5.26 & 5.58 \\
 & \textbf{Waist Width}     & 5.31 & 6.07 & 5.01 & 5.54 & 6.41 & 6.78 \\
\hline
\multirow{4}{*}{\rotatebox{90}{\textbf{A2-A3}}} & \textbf{Clothing Length} & 4.73 & 5.22 & 4.85 & 5.36 & 5.22 & 6.05 \\
 & \textbf{Sleeve Length}   & 6.86 & 7.66 & 5.93 & 6.66 & 6.96 & 7.83 \\
 & \textbf{Shoulder Width}  & 6.09 & 6.60 & 5.29 & 5.73 & 5.65 & 5.98 \\
 & \textbf{Waist Width}     & 5.31 & 5.79 & 5.89 & 6.24 & 6.41 & 6.98 \\
\hline
\end{tabular}
\label{tab: increment_error}
\end{center}
\vspace{2mm}
\small{Note: All values are expressed in percentage (\%).}
\end{table}

Further analysis reveals that clothing length measurements (SMAPE slightly below 5\%) accuracy is higher than shoulder and waist width dimensions. SV-VTON exhibits greater precision differentiating between $A_1$ and $A_2$ styles than between $A_2$ and $A_3$ styles, reflecting the higher practical user demand for distinctions between tight and fit sizing.

In summary, the experimental results verify that the proposed SV-VTON method achieves high precision in multi-dimensional garment sizing, effectively meeting diverse user sizing preferences and significantly enhancing the practicality and generalization capabilities of virtual try-on systems. Although there remains a slight accuracy loss when distinguishing looser sizes, the overall sizing capability aligns well with practical user requirements, providing substantial technical support for real-world virtual try-on tasks.

\subsection{Ablation Study of SV-VTON} 

To thoroughly analyze the effectiveness and impact of input conditions—specifically mask shape and garment image proportion—on the generated results, we designed comparative experiments, as illustrated in Fig.~6(a) and Fig.~6(b). In Fig.~6(a), we investigate the effect of varying the garment image proportion while keeping the mask region fixed. By adjusting the scaling factor of the input garment image, we assess how different proportions influence the try-on results. In contrast, Fig.~6(b) examines the impact of varying the mask shape while maintaining a consistent garment image proportion, allowing us to explore further the mask region's role in try-on image generation. These experiments aim to identify the key input factors that affect the model's ability to differentiate between different garment sizes accurately. 

\begin{figure}[t]
\centerline{\includegraphics[width=\columnwidth]{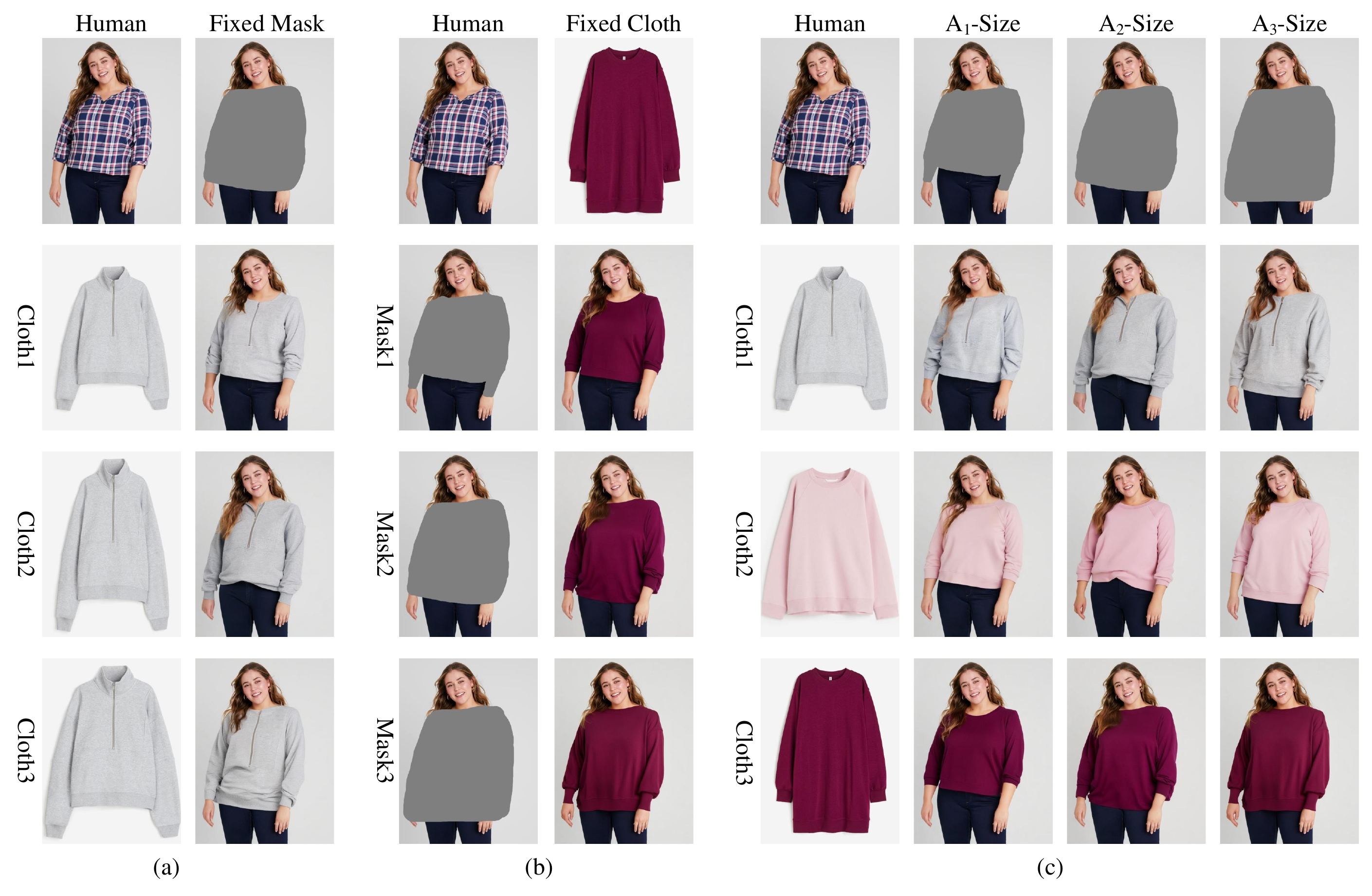}}
\caption{Effect of input conditions on try-on results. 
    (a) Varying the garment image proportion while keeping the mask size fixed to achieve different size styles. 
    (b) Varying the mask size while keeping the garment image proportion fixed to control size variations. 
    (c) Selecting an appropriate combination of mask size and garment image proportion to simulate optimal input conditions for $A_1$-size, $A_2$-size, and $A_3$-size.}
\label{fig6}
\end{figure}

As shown in Fig.~6(a) and Fig.~6(b), we conducted extensive experiments by varying combinations of mask shape and garment image proportion, and experimental results indicate that both garment image proportion and mask shape significantly impact the quality of the generated try-on images. Based on these findings, we determine an optimal combination of input conditions, as shown in Fig.~6(c). The input conditions are as follows: \( A_1 \) uses an \( A_1 \)-size mask with the original garment image proportion (\( C_1 \)); \( A_2 \) uses an \( A_2 \)-size mask and a vertically stretched garment image at 120\% (\( C_2 \)); \( A_3 \) uses an \( A_3 \)-size mask with the garment image stretched to 120\% (\( C_3 \)). Refining these input conditions can give the model more precise constraints, thereby effectively improving its capacity to distinguish between various size styles.

\subsection{Absolute Error Comparison}

Table~\ref{tab:increment_error_mae_rmse} presents the quantitative measurement results of size increments ($A_1$–$A_2$ and $A_2$–$A_3$) for images generated by Reference Net, StableVITON, and DCI-VTON, using two error evaluation metrics: MAE and RMSE. The absolute error of size increments indicates discrepancies between the generated results and the standardized international size increments.

As shown in Fig.~6, the visual differences in length and width between $A_1$-size, $A_2$-size, and $A_3$-size try-on results may not always appear significant. However, the absolute error values in Table~\ref{tab:increment_error_mae_rmse} indicate that the observed differences remain within an acceptable range. Even if some ambiguity exists in the visual appearance, the actual size measurements, adjusted for wrinkles, ensure that the size increments of the generated garment remain within reasonable standardized limits.

As a standard error evaluation metric, MAE effectively quantifies the mean absolute deviation between the measured and reference values. A lower MAE indicates that the generated results are closer to the reference standard. Since international size increments are predetermined fixed values, these serve as the reference standard in our evaluation. As shown in Table~\ref{tab:increment_error_mae_rmse}, the size increments generated by SV-VTON exhibit relatively low mean deviations from the international reference values, with shoulder width and waist width being particularly close to the expected values, demonstrating high accuracy.

\begin{table}[t]
\caption{Increment Error Evaluation (MAE and RMSE) for Diffusion Models}
\begin{center}
\renewcommand{\arraystretch}{1.2} 
\setlength{\tabcolsep}{3pt} 
\begin{tabular}{|c|c|cc|cc|cc|}
\hline
\multicolumn{2}{|c|}{\multirow{2}{*}{\textbf{Increment Error}}} 
& \multicolumn{2}{c|}{\textbf{Reference Net}} 
& \multicolumn{2}{c|}{\textbf{StableVITON}} 
& \multicolumn{2}{c|}{\textbf{DCI-VTON}} \\
\cline{3-8}
\multicolumn{2}{|c|}{} 
& \textbf{MAE}$\downarrow$ & \textbf{RMSE}$\downarrow$ 
& \textbf{MAE} & \textbf{RMSE} 
& \textbf{MAE} & \textbf{RMSE} \\
\hline
\multirow{4}{*}{\rotatebox{90}{\textbf{A1-A2}}} & \textbf{Clothing Length} & 0.208 & 0.319 & 0.219 & 0.339 & 0.308 & 0.409 \\
 & \textbf{Sleeve Length} & 0.180 & 0.260 & 0.171 & 0.255 & 0.354 & 0.378 \\
 & \textbf{Shoulder Width} & 0.156 & 0.228 & 0.146 & 0.201 & 0.203 & 0.248 \\
 & \textbf{Waist Width} & 0.154 & 0.230 & 0.142 & 0.204 & 0.208 & 0.270 \\
\hline
\multirow{4}{*}{\rotatebox{90}{\textbf{A2-A3}}} & \textbf{Clothing Length} & 0.153 & 0.230 & 0.178 & 0.252 & 0.343 & 0.434 \\
 & \textbf{Sleeve Length} & 0.138 & 0.177 & 0.115 & 0.151 & 0.378 & 0.454 \\
 & \textbf{Shoulder Width} & 0.147 & 0.192 & 0.127 & 0.153 & 0.228 & 0.307 \\
 & \textbf{Waist Width} & 0.137 & 0.190 & 0.158 & 0.206 & 0.209 & 0.284 \\
\hline
\end{tabular}
\label{tab:increment_error_mae_rmse}
\end{center}
\vspace{2mm}
\small{Note: All values are expressed in centimeters (cm).}
\end{table}

RMSE, similar to MAE, incorporates a squared computation in its error calculation, making it more sensitive to more significant deviations. Consequently, RMSE values are typically slightly higher than MAE values. However, as seen in Table~\ref{tab:increment_error_mae_rmse}, the difference between MAE and RMSE values is not substantial, indicating that the overall measurement errors are minor and evenly distributed, without extreme outliers. The minor error further confirms the stability and reliability of SV-VTON in generating accurate multi-size images.

\section{Conclusion}

This paper presents SV-VTON, a multi-size VTON framework based on the Diffusion model, enabling personalized virtual fitting across different size styles. We employed the two-stage refined multi-size mask generation strategy, which is introduced for targeted training, overcoming the limitations of existing methods that generate only a single well-fitted try-on result, thereby accommodating diverse user preferences. Moreover, we provide a multi-size selection function, allowing users to freely choose among different size styles—tight-fit ($A_1$), regular-fit ($A_2$), and loose-fit ($A_3$)—while ensuring that garment texture details are well preserved. Additionally, We propose an evaluation metric (EM) to validate that SV-VTON can stably and accurately generate try-on results for multiple sizes based on mainstream Diffusion models, which analyzes the error between the generated garment size increments and international standard size increments. The proposed method expands the potential applications of VTON technology in real-world scenarios.

\section{Limitation and Future Work}

While SV-VTON demonstrates strong overall performance in generating accurate multi-size virtual try-on results, minor deviations still occur in specific generated images. These slight discrepancies may stem from limitations in mask precision and the inherent challenges of maintaining consistent size variations across different garments.  

In future work, we aim to refine the multi-size mask generation process to achieve higher precision and minimize size increment errors. By improving mask detail and enhancing the garment warping mechanism, we seek to enhance the fidelity of generated images, ensuring even more accurate size variations while preserving garment textures and details.

\bibliographystyle{IEEEtran}
\bibliography{ref}

\begin{thebibliography}{10}
\providecommand{\url}[1]{#1}
\csname url@samestyle\endcsname
\providecommand{\newblock}{\relax}
\providecommand{\bibinfo}[2]{#2}
\providecommand{\BIBentrySTDinterwordspacing}{\spaceskip=0pt\relax}
\providecommand{\BIBentryALTinterwordstretchfactor}{4}
\providecommand{\BIBentryALTinterwordspacing}{\spaceskip=\fontdimen2\font plus
\BIBentryALTinterwordstretchfactor\fontdimen3\font minus \fontdimen4\font\relax}
\providecommand{\BIBforeignlanguage}[2]{{%
\expandafter\ifx\csname l@#1\endcsname\relax
\typeout{** WARNING: IEEEtran.bst: No hyphenation pattern has been}%
\typeout{** loaded for the language `#1'. Using the pattern for}%
\typeout{** the default language instead.}%
\else
\language=\csname l@#1\endcsname
\fi
#2}}
\providecommand{\BIBdecl}{\relax}
\BIBdecl

\bibitem{1}
G.~Eason, B.~Noble, and I.~N. Sneddon, ``On certain integrals of lipschitz-hankel type involving products of bessel functions,'' \emph{Philosophical Transactions of the Royal Society of London. Series A, Mathematical and Physical Sciences}, vol. 247, no. 935, pp. 529--551, 1955.

\bibitem{7}
S.~Choi, S.~Park, M.~Lee, and J.~Choo, ``Viton-hd: High-resolution virtual try-on via misalignment-aware normalization,'' in \emph{Proceedings of the IEEE/CVF conference on computer vision and pattern recognition}, 2021, pp. 14\,131--14\,140.

\bibitem{10}
X.~Han, Z.~Wu, Z.~Wu, R.~Yu, and L.~S. Davis, ``Viton: An image-based virtual try-on network,'' in \emph{Proceedings of the IEEE conference on computer vision and pattern recognition}, 2018, pp. 7543--7552.

\bibitem{2}
X.~Han, X.~Hu, W.~Huang, and M.~R. Scott, ``Clothflow: A flow-based model for clothed person generation,'' in \emph{Proceedings of the IEEE/CVF international conference on computer vision}, 2019, pp. 10\,471--10\,480.

\bibitem{18}
S.~Lee, G.~Gu, S.~Park, S.~Choi, and J.~Choo, ``High-resolution virtual try-on with misalignment and occlusion-handled conditions,'' in \emph{European Conference on Computer Vision}.\hskip 1em plus 0.5em minus 0.4em\relax Springer, 2022, pp. 204--219.

\bibitem{21}
T.~Issenhuth, J.~Mary, and C.~Calauzenes, ``Do not mask what you do not need to mask: a parser-free virtual try-on,'' in \emph{Computer Vision--ECCV 2020: 16th European Conference, Glasgow, UK, August 23--28, 2020, Proceedings, Part XX 16}.\hskip 1em plus 0.5em minus 0.4em\relax Springer, 2020, pp. 619--635.

\bibitem{3}
C.-Y. Chen, Y.-C. Chen, H.-H. Shuai, and W.-H. Cheng, ``Size does matter: Size-aware virtual try-on via clothing-oriented transformation try-on network,'' in \emph{Proceedings of the IEEE/CVF international conference on computer vision}, 2023, pp. 7513--7522.

\bibitem{4}
S.~Zhang, M.~Ni, S.~Chen, L.~Wang, W.~Ding, and Y.~Liu, ``A two-stage personalized virtual try-on framework with shape control and texture guidance,'' \emph{IEEE Transactions on Multimedia}, 2024.

\bibitem{6}
N.~Fang, L.~Qiu, S.~Zhang, Z.~Wang, and K.~Hu, ``Pg-vton: A novel image-based virtual try-on method via progressive inference paradigm,'' \emph{IEEE Transactions on Multimedia}, vol.~26, pp. 6595--6608, 2024.

\bibitem{8}
A.~Cui, J.~Mahajan, V.~Shah, P.~Gomathinayagam, C.~Liu, and S.~Lazebnik, ``Street tryon: Learning in-the-wild virtual try-on from unpaired person images,'' in \emph{Proceedings of the IEEE/CVF Conference on Computer Vision and Pattern Recognition}, 2024, pp. 8235--8239.

\bibitem{24}
J.~Ho, A.~Jain, and P.~Abbeel, ``Denoising diffusion probabilistic models,'' \emph{Advances in neural information processing systems}, vol.~33, pp. 6840--6851, 2020.

\bibitem{9}
H.~Dong, X.~Liang, X.~Shen, B.~Wang, H.~Lai, J.~Zhu, Z.~Hu, and J.~Yin, ``Towards multi-pose guided virtual try-on network,'' in \emph{Proceedings of the IEEE/CVF international conference on computer vision}, 2019, pp. 9026--9035.

\bibitem{11}
Z.~Liu, P.~Luo, S.~Qiu, X.~Wang, and X.~Tang, ``Deepfashion: Powering robust clothes recognition and retrieval with rich annotations,'' in \emph{Proceedings of the IEEE conference on computer vision and pattern recognition}, 2016, pp. 1096--1104.

\bibitem{12}
D.~Morelli, M.~Fincato, M.~Cornia, F.~Landi, F.~Cesari, and R.~Cucchiara, ``Dress code: High-resolution multi-category virtual try-on,'' in \emph{Proceedings of the IEEE/CVF conference on computer vision and pattern recognition}, 2022, pp. 2231--2235.

\bibitem{48}
X.~Qin, Z.~Zhang, C.~Huang, M.~Dehghan, O.~R. Zaiane, and M.~Jagersand, ``U2-net: Going deeper with nested u-structure for salient object detection,'' \emph{Pattern recognition}, vol. 106, p. 107404, 2020.

\bibitem{19}
Z.~Xie, Z.~Huang, X.~Dong, F.~Zhao, H.~Dong, X.~Zhang, F.~Zhu, and X.~Liang, ``Gp-vton: Towards general purpose virtual try-on via collaborative local-flow global-parsing learning,'' in \emph{Proceedings of the IEEE/CVF Conference on Computer Vision and Pattern Recognition}, 2023, pp. 23\,550--23\,559.

\bibitem{20}
S.~Bai, H.~Zhou, Z.~Li, C.~Zhou, and H.~Yang, ``Single stage virtual try-on via deformable attention flows,'' in \emph{European Conference on Computer Vision}.\hskip 1em plus 0.5em minus 0.4em\relax Springer, 2022, pp. 409--425.

\bibitem{16}
T.~Chong, I.-C. Shen, N.~Umetani, and T.~Igarashi, ``Per garment capture and synthesis for real-time virtual try-on,'' in \emph{The 34th Annual ACM Symposium on User Interface Software and Technology}, 2021, pp. 457--469.

\bibitem{17}
Y.~Ge, Y.~Song, R.~Zhang, C.~Ge, W.~Liu, and P.~Luo, ``Parser-free virtual try-on via distilling appearance flows,'' in \emph{Proceedings of the IEEE/CVF conference on computer vision and pattern recognition}, 2021, pp. 8485--8493.

\bibitem{13}
Z.~Yang, J.~Chen, Y.~Shi, H.~Li, T.~Chen, and L.~Lin, ``Occlumix: Towards de-occlusion virtual try-on by semantically-guided mixup,'' \emph{IEEE Transactions on Multimedia}, vol.~25, pp. 1477--1488, 2023.

\bibitem{14}
S.~Zhang, X.~Han, W.~Zhang, X.~Lan, H.~Yao, and Q.~Huang, ``Limb-aware virtual try-on network with progressive clothing warping,'' \emph{IEEE Transactions on Multimedia}, vol.~26, pp. 1731--1746, 2023.

\bibitem{15}
J.~Wang, P.~Liu, J.~Liu, and W.~Xu, ``Text-guided eyeglasses manipulation with spatial constraints,'' \emph{IEEE Transactions on Multimedia}, vol.~26, pp. 4375--4388, 2023.

\bibitem{5}
L.~Hu, ``Animate anyone: Consistent and controllable image-to-video synthesis for character animation,'' in \emph{Proceedings of the IEEE/CVF Conference on Computer Vision and Pattern Recognition}, 2024, pp. 8153--8163.

\bibitem{22}
L.~Zhu, D.~Yang, T.~Zhu, F.~Reda, W.~Chan, C.~Saharia, M.~Norouzi, and I.~Kemelmacher-Shlizerman, ``Tryondiffusion: A tale of two unets,'' in \emph{Proceedings of the IEEE/CVF Conference on Computer Vision and Pattern Recognition}, 2023, pp. 4606--4615.

\bibitem{25}
B.~Yang, S.~Gu, B.~Zhang, T.~Zhang, X.~Chen, X.~Sun, D.~Chen, and F.~Wen, ``Paint by example: Exemplar-based image editing with diffusion models,'' in \emph{Proceedings of the IEEE/CVF conference on computer vision and pattern recognition}, 2023, pp. 18\,381--18\,391.

\bibitem{26}
R.~Rombach, A.~Blattmann, D.~Lorenz, P.~Esser, and B.~Ommer, ``High-resolution image synthesis with latent diffusion models,'' in \emph{Proceedings of the IEEE/CVF conference on computer vision and pattern recognition}, 2022, pp. 10\,684--10\,695.

\bibitem{28}
J.~Gou, S.~Sun, J.~Zhang, J.~Si, C.~Qian, and L.~Zhang, ``Taming the power of diffusion models for high-quality virtual try-on with appearance flow,'' in \emph{Proceedings of the 31st ACM International Conference on Multimedia}, 2023, pp. 7599--7607.

\bibitem{27}
D.~Morelli, A.~Baldrati, G.~Cartella, M.~Cornia, M.~Bertini, and R.~Cucchiara, ``Ladi-vton: Latent diffusion textual-inversion enhanced virtual try-on,'' in \emph{Proceedings of the 31st ACM International Conference on Multimedia}, 2023, pp. 8580--8589.

\bibitem{45}
J.~Kim, G.~Gu, M.~Park, S.~Park, and J.~Choo, ``Stableviton: Learning semantic correspondence with latent diffusion model for virtual try-on,'' in \emph{Proceedings of the IEEE/CVF conference on computer vision and pattern recognition}, 2024, pp. 8176--8185.

\bibitem{31}
G.~Kim, T.~Kwon, and J.~C. Ye, ``Diffusionclip: Text-guided diffusion models for robust image manipulation,'' in \emph{Proceedings of the IEEE/CVF conference on computer vision and pattern recognition}, 2022, pp. 2426--2435.

\bibitem{32}
L.~Zhu, Y.~Li, N.~Liu, H.~Peng, D.~Yang, and I.~Kemelmacher-Shlizerman, ``M\&m vto: Multi-garment virtual try-on and editing,'' in \emph{Proceedings of the IEEE/CVF Conference on Computer Vision and Pattern Recognition}, 2024, pp. 1346--1356.

\bibitem{33}
S.~Ning, D.~Wang, Y.~Qin, Z.~Jin, B.~Wang, and X.~Han, ``Picture: Photorealistic virtual try-on from unconstrained designs,'' in \emph{Proceedings of the IEEE/CVF conference on computer vision and pattern recognition}, 2024, pp. 6976--6985.

\bibitem{34}
X.~Yang, C.~Ding, Z.~Hong, J.~Huang, J.~Tao, and X.~Xu, ``Texture-preserving diffusion models for high-fidelity virtual try-on,'' in \emph{Proceedings of the IEEE/CVF conference on computer vision and pattern recognition}, 2024, pp. 7017--7026.

\bibitem{42}
M.~R. Minar, T.~T. Tuan, H.~Ahn, P.~Rosin, and Y.-K. Lai, ``Cp-vton+: Clothing shape and texture preserving image-based virtual try-on,'' in \emph{CVPR workshops}, vol.~3, 2020, pp. 10--14.

\bibitem{29}
L.~Zhang, A.~Rao, and M.~Agrawala, ``Adding conditional control to text-to-image diffusion models,'' in \emph{Proceedings of the IEEE/CVF international conference on computer vision}, 2023, pp. 3836--3847.

\bibitem{38}
A.~Hore and D.~Ziou, ``Image quality metrics: Psnr vs. ssim,'' in \emph{2010 20th international conference on pattern recognition}.\hskip 1em plus 0.5em minus 0.4em\relax IEEE, 2010, pp. 2366--2369.

\bibitem{49}
M.~Heusel, H.~Ramsauer, T.~Unterthiner, B.~Nessler, and S.~Hochreiter, ``Gans trained by a two time-scale update rule converge to a local nash equilibrium,'' \emph{Advances in neural information processing systems}, vol.~30, 2017.

\bibitem{50}
M.~Bi{\'n}kowski, D.~J. Sutherland, M.~Arbel, and A.~Gretton, ``Demystifying mmd gans,'' \emph{arXiv preprint arXiv:1801.01401}, 2018.

\bibitem{51}
Z.~Wang, E.~P. Simoncelli, and A.~C. Bovik, ``Multiscale structural similarity for image quality assessment,'' in \emph{The Thrity-Seventh Asilomar Conference on Signals, Systems \& Computers, 2003}, vol.~2.\hskip 1em plus 0.5em minus 0.4em\relax Ieee, 2003, pp. 1398--1402.

\bibitem{39}
R.~Zhang, P.~Isola, A.~A. Efros, E.~Shechtman, and O.~Wang, ``The unreasonable effectiveness of deep features as a perceptual metric,'' in \emph{Proceedings of the IEEE conference on computer vision and pattern recognition}, 2018, pp. 586--595.

\bibitem{44}
J.~Hessel, A.~Holtzman, M.~Forbes, R.~L. Bras, and Y.~Choi, ``Clipscore: A reference-free evaluation metric for image captioning,'' \emph{arXiv preprint arXiv:2104.08718}, 2021.

\bibitem{40}
Z.~Wang, A.~C. Bovik, H.~R. Sheikh, and E.~P. Simoncelli, ``Image quality assessment: from error visibility to structural similarity,'' \emph{IEEE transactions on image processing}, vol.~13, no.~4, pp. 600--612, 2004.

\bibitem{41}
A.~Cui, D.~McKee, and S.~Lazebnik, ``Dressing in order: Recurrent person image generation for pose transfer, virtual try-on and outfit editing,'' in \emph{Proceedings of the IEEE/CVF international conference on computer vision}, 2021, pp. 14\,638--14\,647.

\bibitem{47}
D.~Song, X.~Zhang, J.~Zeng, P.~Zhan, Q.~Chen, W.~Luo, and A.-A. Liu, ``Better fit: Accommodate variations in clothing types for virtual try-on,'' \emph{arXiv preprint arXiv:2403.08453}, 2024.

\bibitem{52}
S.~Yang, T.~Wu, S.~Shi, S.~Lao, Y.~Gong, M.~Cao, J.~Wang, and Y.~Yang, ``Maniqa: Multi-dimension attention network for no-reference image quality assessment,'' in \emph{Proceedings of the IEEE/CVF conference on computer vision and pattern recognition}, 2022, pp. 1191--1200.

\bibitem{53}
J.~Ke, Q.~Wang, Y.~Wang, P.~Milanfar, and F.~Yang, ``Musiq: Multi-scale image quality transformer,'' in \emph{Proceedings of the IEEE/CVF international conference on computer vision}, 2021, pp. 5148--5157.

\bibitem{36}
Z.~Zhang, H.~Fu, H.~Dai, J.~Shen, Y.~Pang, and L.~Shao, ``Et-net: A generic edge-attention guidance network for medical image segmentation,'' in \emph{Medical Image Computing and Computer Assisted Intervention--MICCAI 2019: 22nd International Conference, Shenzhen, China, October 13--17, 2019, Proceedings, Part I 22}.\hskip 1em plus 0.5em minus 0.4em\relax Springer, 2019, pp. 442--450.

\bibitem{37}
H.~Chen, X.~Qi, L.~Yu, Q.~Dou, J.~Qin, and P.-A. Heng, ``Dcan: Deep contour-aware networks for object instance segmentation from histology images,'' \emph{Medical image analysis}, vol.~36, pp. 135--146, 2017.

\end{thebibliography}

\begin{IEEEbiography}[{\includegraphics[width=1in,height=1.25in,clip,keepaspectratio]{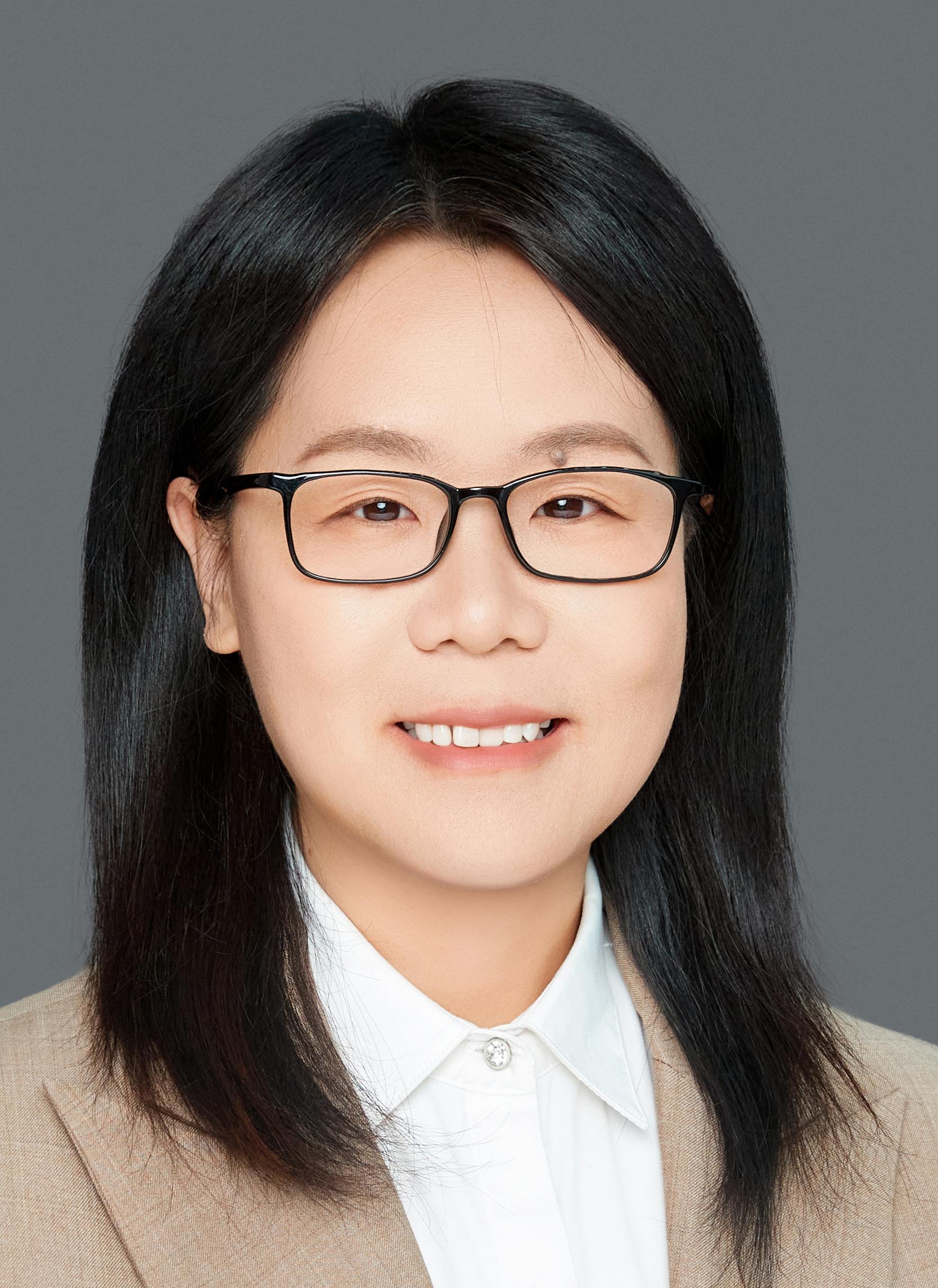}}]{Shufang Zhang}
is currently an associate professor in the School of Electrical Automation and Information Engineering, Tianjin University, Tianjin, China. She received her M.S. and Ph.D. degrees from Tianjin University in 2004 and 2007, respectively. Her research interests include deep learning, virtual try-on, and outfit recommendation.
\end{IEEEbiography}

\begin{IEEEbiography}[{\includegraphics[width=1in,height=1.25in,clip,keepaspectratio]{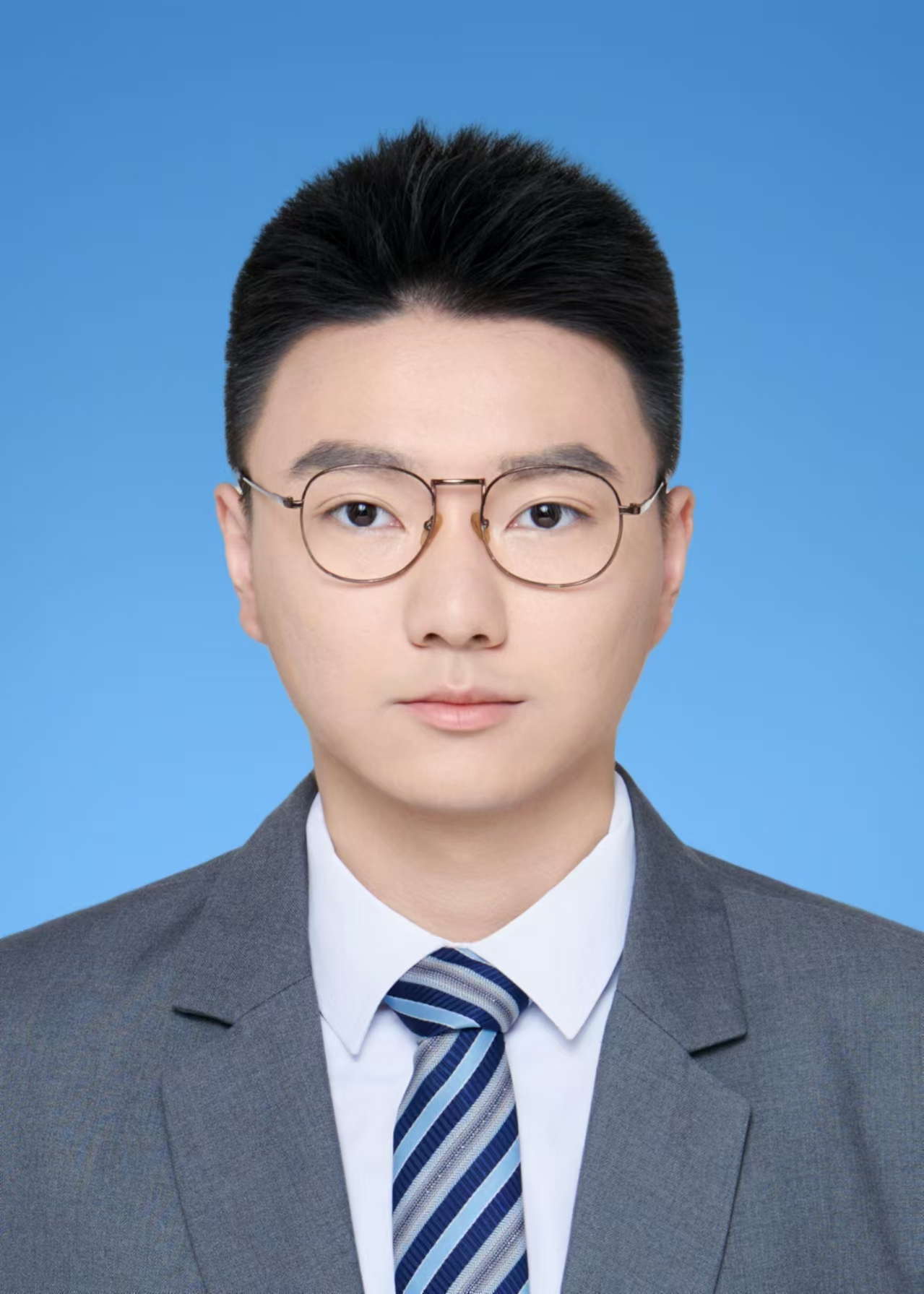}}]{Hang Qian}
received a B.S. degree in the School of Electrical Automation and Information Engineering from Tianjin University in 2023, Tianjin, China. He is currently working toward his Master's degree in the School of Electrical and Information Engineering at Tianjin University, Tianjin, China. His research interests mainly focus on virtual try-on based on image.
\end{IEEEbiography}

\begin{IEEEbiography}[{\includegraphics[width=1in,height=1.25in,clip,keepaspectratio]{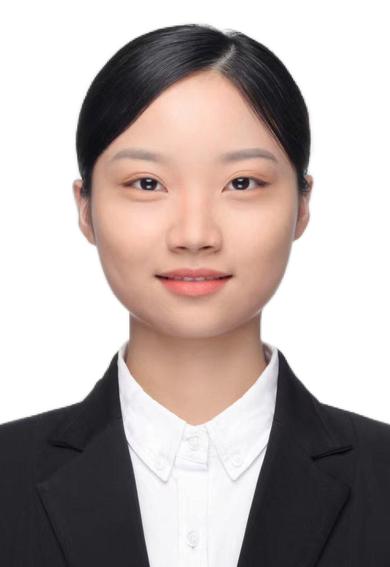}}]{Minxue Ni}
received a B.S. degree in the School of Electrical Automation and Information Engineering from Tianjin University in 2022, Tianjin, China. She is currently working toward her Master's degree at Tianjin University. Her research interests mainly include computer vision for virtual try-on.
\end{IEEEbiography}

\begin{IEEEbiography}[{\includegraphics[width=1in,height=1.25in,clip,keepaspectratio]{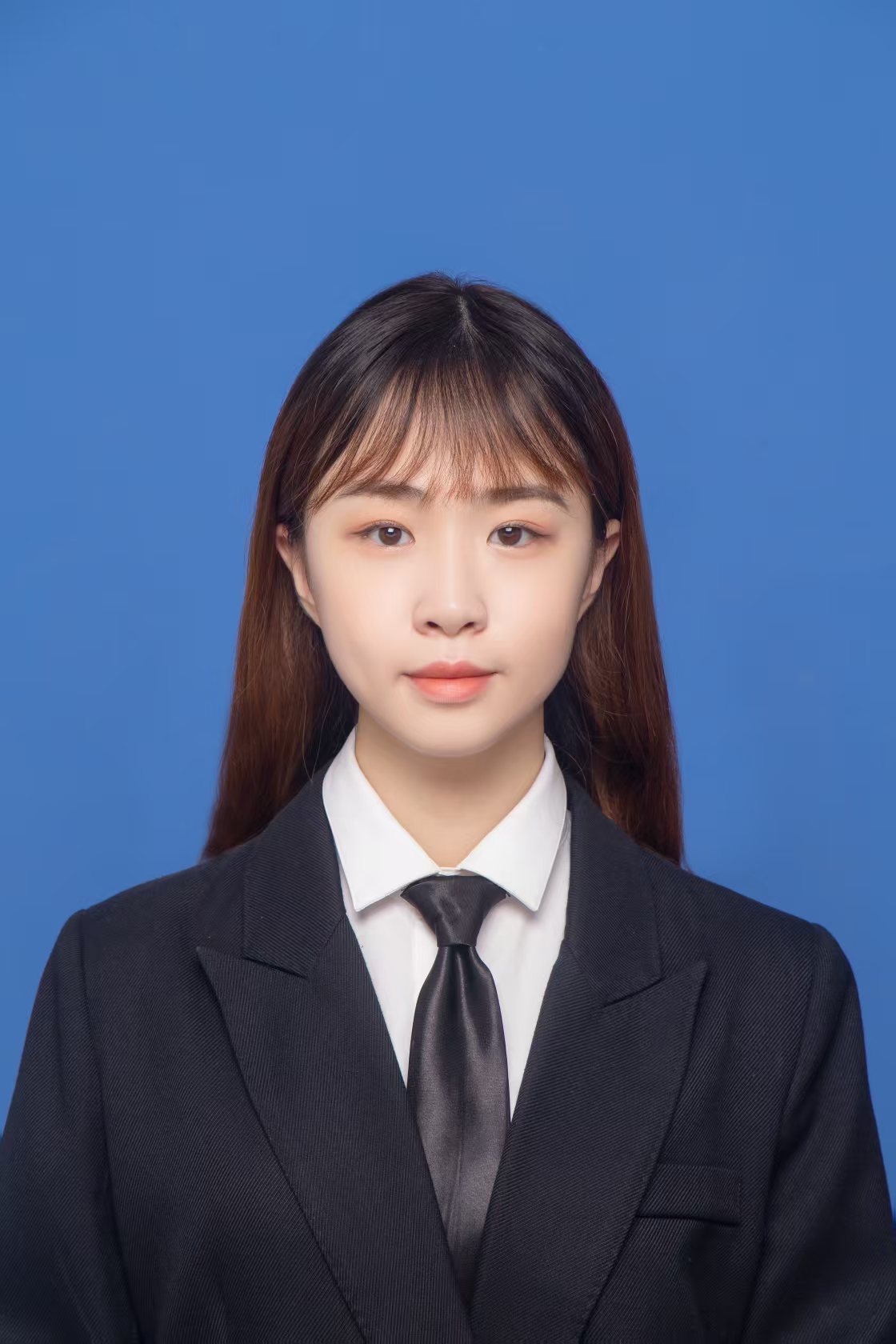}}]{Yaxuan Li}
received a B.S. degree in 2024 from Tianjin University, Tianjin, China. She is currently working toward her Master's degree at Tianjin University. Her current research is virtual try-on.
\end{IEEEbiography}

\begin{IEEEbiography}[{\includegraphics[width=1in,height=1.25in,clip,keepaspectratio]{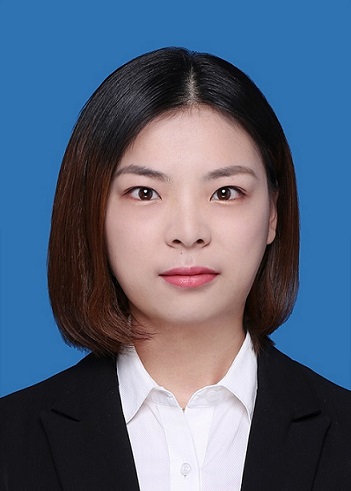}}]{Wenxin Ding}
received the B.S. and M.S. degrees in the School of Electrical Automation and Information Engineering from Tianjin University in 2016 and 2019, respectively. She is currently pursuing her Ph.D. in the School of Electrical and Information Engineering, Tianjin University, Tianjin, China. Her research mainly focuses on outfit recommendation and virtual try-on.
\end{IEEEbiography}

\begin{IEEEbiography}[{\includegraphics[width=1in,height=1.25in,clip,keepaspectratio]{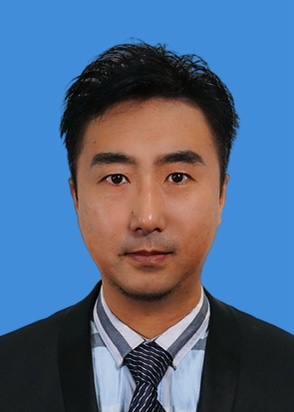}}]{Jun Liu}
is currently an associate professor and graduate supervisor at the School of Humanities and Arts, Tianjin University, China. His research interests include visual art design and image effect evaluation.
\end{IEEEbiography}

\end{document}